\documentclass[a4paper,fleqn,usenatbib]{mnras}

\usepackage{newtxtext,newtxmath}

\usepackage[normalem]{ ulem }
\usepackage{soul}
\usepackage[T1]{fontenc}
\usepackage{ae,aecompl}
\usepackage{xcolor}

\usepackage{graphicx}	
\usepackage{amsmath}	
\usepackage{amssymb}	
\usepackage{enumitem}
\setlist[enumerate,1]{start=0}

\definecolor{orange}{rgb}{.9,.6,0}
\definecolor{skyblue}{rgb}{.35,.7,.9}
\definecolor{bluishgreen}{rgb}{0,0.6,0.5}
\definecolor{yellow}{rgb}{0.95,0.9,0.25}
\definecolor{blue}{rgb}{0,0.45,0.7}
\definecolor{vermillon}{rgb}{0.8,0.4,0}
\definecolor{reddishpurple}{rgb}{0.8,0.6,0.7}

\title[Mode Classification in Fast-Rotating Stars]
{Mode Classification in Fast-Rotating Stars using a Convolutional Neural Network: Model-Based Regular Patterns in $\delta$ Scuti Stars}

\author[G.M. Mirouh]{
Giovanni M. Mirouh,$^{1,2,3}$\thanks{Email: gmirouh@surrey.ac.uk}
George C. Angelou$^{2}$,
Daniel R. Reese$^{4}$,
and Guglielmo Costa$^{1}$
\\
$^{1}$SISSA, Via Bonomea 265, 34136 Trieste, Italy\\
$^{2}$Max-Planck-Institut f\"{u}r Astrophysik, Karl-Schwarzschild-Str. 1, 85748, Garching, Germany \\
$^{3}$Astrophysics Research Group, Faculty of Engineering and Physical Sciences, University of Surrey, Guildford, GU2 7XH, United Kingdom\\
$^{4}$LESIA, Observatoire de Paris, Universit{\'e} PSL, Sorbonne Universit{\'e}, Univ. Paris Diderot, Sorbonne Paris Cit{\'e}, 5 place Jules Janssen, 92195 Meudon, France
\\
}  

\date{Accepted XXX. Received YYY; in original form ZZZ}

\pubyear{2018}

\begin{document}
\label{firstpage}
\pagerange{\pageref{firstpage}--\pageref{lastpage}}
\maketitle

\begin{abstract}
Oscillation modes in fast-rotating stars can be split into several subclasses,
  each with their own properties.  To date, seismology of these stars cannot
  rely on regular pattern analysis and scaling relations. However, recently
  there has been the promising discovery of large separations observed in
  spectra of fast-rotating $\delta$ Scuti stars: they were attributed to the
  island-mode subclass, and linked to the stellar mean density through a
  scaling law.  In this work, we investigate the relevance of this scaling
  relation by computing models of fast-rotating stars and their oscillation
  spectra.  In order to sort the thousands of oscillation modes thus obtained,
  we train a convolutional neural network isolating the island modes with 96\%
  accuracy.  Arguing that the observed large separation is systematically
  smaller than the asymptotic one, we retrieve the observational $\Delta\nu -
  \overline{\rho}$ scaling law.  This relation will be used to drive forward
  modelling efforts, and is a first step towards mode identification and
  inversions for fast-rotating stars.

\end{abstract}

\begin{keywords}
stars: oscillations -- stars: rotation -- variables: $\delta$ Scuti
\end{keywords}

\section{Introduction}
\subsection{Fast-rotating stars}
Through space missions such as MOST, CoRoT and Kepler, asteroseismology has
proven to be the most powerful tool by which to probe stellar interiors.  Most
of the results obtained through this technique rely on the regular patterns the
mode frequencies follow, that can readily be linked to the stellar fundamental
parameters through scaling laws. However, this technique works for
slowly-rotating solar-like oscillators, where mode identification is possible.
Deeper insight can be gained through the analysis of rotational splittings,
automated inferences or inversion techniques.  In fast-rotating stars, such
identification is not so easy: while the centrifugal force distorts the stellar
geometry, the Coriolis force complicates mode geometries such that they can no
longer be described in terms of simple spherical harmonics.  Currently, it is
standard to use a linear combination of spherical harmonics as a basis to
describe the modes, but this prevents mode identification in terms of the
classical quantum numbers $(n,\ell,m)$.

\subsection{Classes of modes}
Theoretical works show that pressure modes in fast-rotating stars can be sorted
in different categories.  \citet{LG09} split pressure modes in fast-rotating
stars in four sub-classes: 2-period island modes, 6-period island modes,
whispering gallery modes and chaotic modes.  Each of these subclasses has its
own regular spacing in frequency.  In measured spectra however, those spacings
cannot readily be distinguished and the associated information on the stellar
structure cannot easily be retrieved. \\ \citet{AGH2015,AGH2017} analyzed a
sample of ten stars, and identified regular patterns in the their
high-frequency spectra. They found a large frequency separation that they
attribute to 2-period island modes, and linked this separation to the stellar
mean density through a scaling law, as predicted by \citet{RLR08}.  In this
work, we provide further validation by exploring theoretical models and their
oscillations to test the proposed scaling law.

\section{Method}

\subsection{Codes}
We compute rotating star models using the two-dimensional structure code ESTER
(\citet{REL13}, \citet{RELP2016}). Adiabatic oscillations of these models are
calculated using TOP (\citet{RTMJSM09}, Reese et al. in prep.). The geometry
these codes use rely on the definition of a pseudoradius $\zeta$, that goes
from 0 at the center of the star to 1 at the distorted stellar surface
\citep[see ][]{BGM98}.  The radial grid is split into eight
Gauss-Lobatto-Chebyshev subgrids of 30 points, while the latitudinal components
are projected on 24 spherical harmonics for the structure and 40 spherical
harmonics for the oscillations. This resolution leads to 48,000 modes in the
whole spectrum that are of potential interest.

\subsection{Convolutional Neural Network Classifier} 
\subsubsection {Network Architecture}
\label{sec:cnn}
Previously, mode sorting was performed manually through plotting and visually
inspecting the TOP eigensolutions.  Given that this task is essentially an
image classification problem, the process can be automated through the use of a
convolutional neural network (CNN).  In the last few years machine learning
algorithms have become more widely adopted in stellar astrophysics
\citep{2016ApJ...830...31B, 2016MNRAS.461.4206V, 2017ApJ...839..116A} including
CNNs \citep{2017MNRAS.469.4578H}.

We utilised Google's Tensorflow libraries \citep{tensorflow2015} to create a
seven layer, 2D-convolutional network for the purposes of classifying the TOP
oscillation modes.  We employ a network architecture comprising two
convolution, two pooling and two fully connected layers supplemented by an
output layer.  A rectified linear unit activation function is applied to the
convolution and fully-connected layers.  Max-pooling is used to downsample the
convolutions and drop out is employed as form of regularization for the
fully-connected layers.  A softmax function is used to activate the output
layer for the purpose of assigning class probabilities. 

Before applying the algorithm to mode classification in rapidly-rotating stars
we conducted several tests of our CNN.  We validated our network on the MNIST
database of handwritten digits with 99.3\% accuracy \citep{MNIST}.  As per
\citet{2017MNRAS.469.4578H}, we classified the evolutionary phase  of Kepler
giants in the  \citet{2016A&A...588A..87V} sample with 98\% accuracy.

\subsubsection{Training and development set for mode classification}
\begin{figure}
    \centering
    \includegraphics[width=.5\columnwidth]{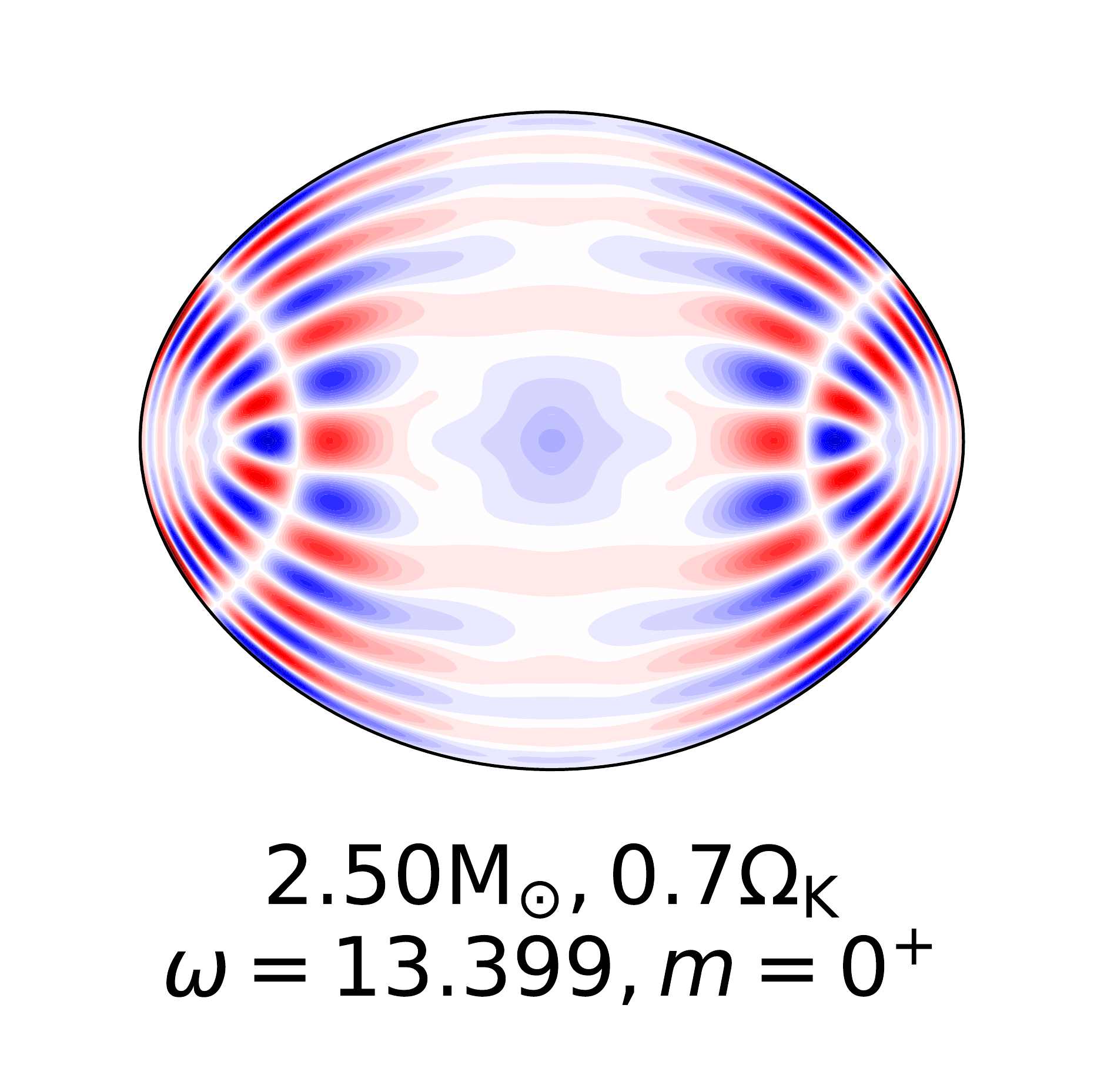} 
    \includegraphics[width=.4\columnwidth]{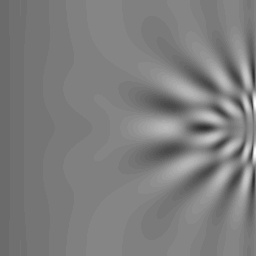}

    \caption{Island pressure mode at
    $(\widetilde{n},\widetilde{\ell},m)=(16,1,0)$, used in the training set.
    Left: meridional cross-section of the Eulerian pressure perturbation
    divided by the square root of the background density. Right: Same quantity
    plotted as a function of the pseudoradius $\zeta$ and the colatitude
    $\theta$.} 
    \label{fig:example}
\end{figure}

To visualize the oscillations through the model, we represent the ratio of the
Eulerian pressure perturbation to the square root of the background density;
this quantity brings out surface variations. For visual inspection purposes, we
plot this quantity along a meridional cross-section.  We feed the algorithm
128x128 (pixels) grey scale images of the same quantity, plotted in the
pseudoradius-colatitude $(\zeta,\theta)$ plane, for $\theta$ going from 0 to
$\pi$.  Figure~\ref{fig:example} shows these two representations for a given
oscillation.

To train the algorithm, we classify by eye 4300 modes divided into seven classes: 
    (i) spurious modes,
    (ii) rosette g modes,
    (iii) subcritical g modes,
    (iv) g modes with some envelope extent,
    (v) whispering gallery p modes,
    (vi) island p modes of period 2,
    (vii) other p modes.

We only keep modes fitting the canonical description of the modes given by
\citet{LG09}, omitting mixed modes and modes resulting from avoided crossings.
Although we do not exploit the whole set of available mode types, the sorted
classes are sufficient for our purposes.

The 4300 images with known truth labels were divided into a training set (80\%)
and a development set (20\%).  We performed a 10-fold validation test which
yielded a mean accuracy of 96\% on randomly selected development sets.  During
the training process we optimized for 3000 iterations after which there was no
significant improvement to our loss metric.  We found that for the current
application, the CNN was most responsive to the Adam optimizer.  As our aim is
to identify 2-period island modes, an accuracy of 96\% was deemed satisfactory.
False positives in this category could be discarded by eye.  We note that there
is scope to optimize the CNN performance further, and we will continue to do so
in future work.

\section{Regular patterns in the island-mode spectrum}
\subsection{Theoretical models and oscillations}
We consider two series of ESTER models of 2.5 $M_\odot$ main-sequence stars: a
series of ZAMS models for increasing rotation velocities and a series of models
rotating at 70\% of their Keplerian rotation rate with varying core hydrogen
abundance $X_c$ to mimic main-sequence evolution.

Each of these models is computed for metallicities $Z=0.02$ and $Z=0.01$. These
models cover the rotation and core abundance parameter space accessible for a
$2.5 M_\odot\ \delta$ Scuti star. Using the TOP code, we
compute the adiabatic, even (symmetric with respect to the underlying ray path)
axisymmetric ($m=0$) oscillations of these models.  We consider a high
frequency interval, where we expect mostly pressure modes, and find about 500
modes in the chosen range for each model. These modes are then fed into the
convolutional neural network described in section~\ref{sec:cnn}, keeping the
modes identified with a probability of 95\%.

\subsection{Comparison with the observations}
Island modes of period 2 are known to be the rotating counterpart of low-degree
pressure modes \citep{Pasek2012}.  In order to identify and study these modes,
we use the description from \citet{RTMJSM09}: each island mode is described
using three quantum numbers, $\widetilde{n}, \widetilde{\ell},m$.  The first
two quantities are illustrated on figure~\ref{fig:example}: $\widetilde{n}$ is
the number of nodes along the wave train from the two points where it reaches
the surface, while $\widetilde{\ell}$ being the number of parallels (i.e. the
number of nodal lines parallel to the equator) from pole to equator.  The
azimuthal order $m$ does not change definition with respect to the non-rotating
case, and is equal to the periodicity in the azimuthal direction.

\citet{LG09} described the island modes in a polytropic rotating fluid, and
highlighted regular patterns in their frequency spectra.  Successive modes at a
given value of $\widetilde{\ell}$ are separated by a frequency distance that
reaches an asymptotic value for high-frequency modes: this value is usually
called the large separation. 

\begin{figure}
    \centering
    \includegraphics[width=\columnwidth]{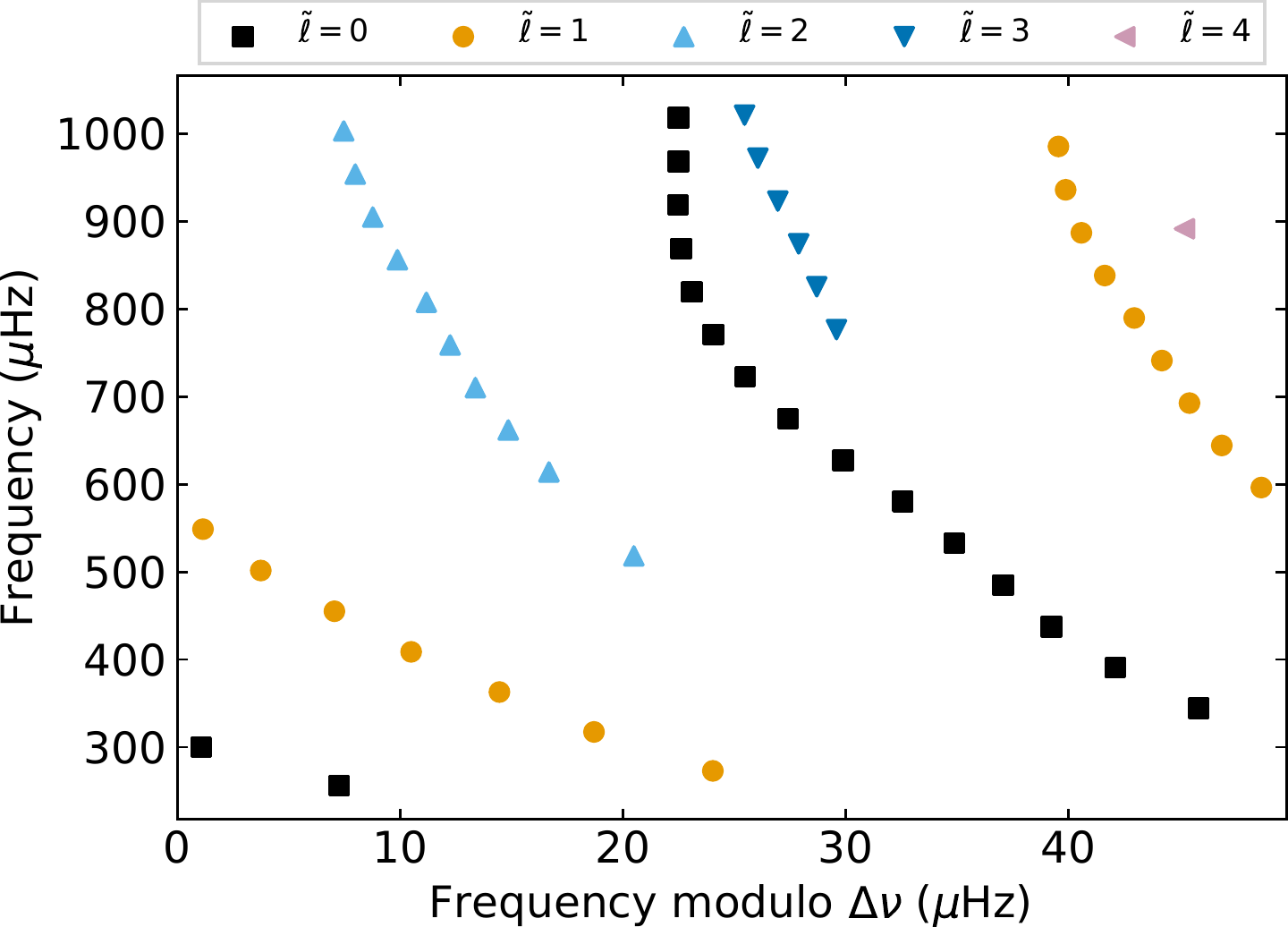}
    \caption{Identification of island modes obtained through our neural network
    for a $Z=0.02$, 2.5 $M_\odot$ model rotating at 70\% of its breakup
    velocity, with 60\% of its initial hydrogen abundance in the core.  Here,
    $\Delta\nu=49.78\mu$Hz in the asymptotic regime.}
    \label{fig:pattern}
\end{figure}

Figure~\ref{fig:pattern} shows an \'echelle diagram for one of the models we
examined. Each ridge corresponds to a different value of $\widetilde{\ell}$
(obtained here through visual inspection). A regular separation appears and
stabilizes when reaching the high-frequency asymptotic regime, thus confirming
the result obtained on polytropic models by \citet{LG09}.

\citet{AGH2015,AGH2017} analyzed observed oscillation spectra for several
rotating $\delta$ Scuti stars. They identified statistically significant
patterns in the frequency spectra.  They were then able to correlate these
large separations with the mean density of the star through a power law that
reads:
\begin{equation}
\frac{\overline{\rho}}{\overline{\rho}_\odot} = 
                   1.55^{+1.07}_{-0.68} \left( \frac{\Delta\nu}{\Delta\nu_\odot} \right)^{2.035\pm 0.095},
\label{eq:agh}
\end{equation}
with $\overline{\rho}_\odot = 1.41$ g/cm$^3$ and $\Delta\nu_\odot = 134.8\mu Hz$ \citep{KBCD08}

\begin{figure}
    \centering
    \includegraphics[width=\columnwidth]{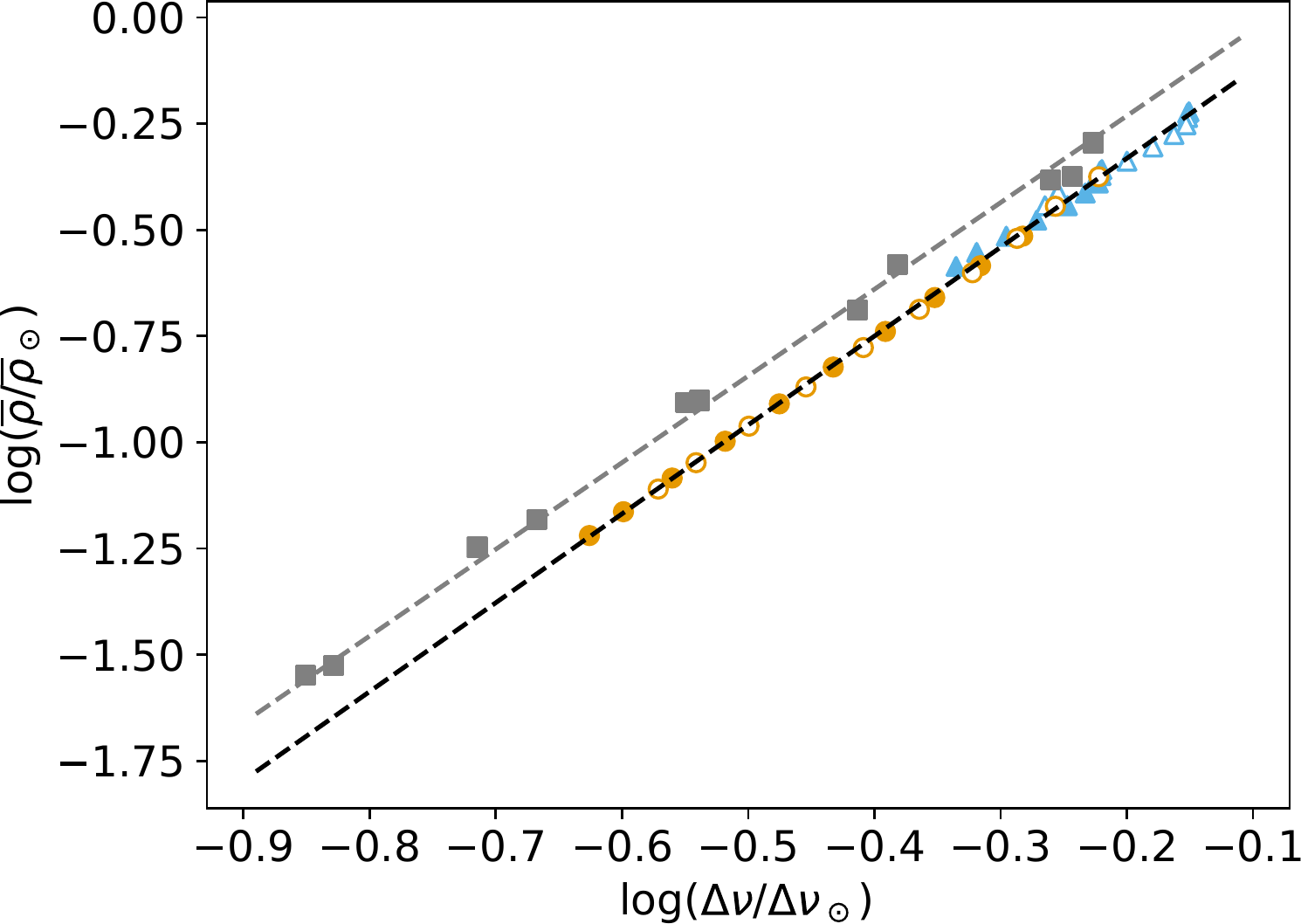}
    \caption{Stellar mean density as a function of the island-mode large separation computed in the asymptotic regime.
    Coloured data points are from our calculations. Empty symbols are for $Z=0.01$ and filled symbols for $Z=0.02$. 
    Blue triangles correspond to varying rotational velocities, orange circles 
    to varying hydrogen core abundances. The black dotted line is our fit using all our models.
    The grey squares are the data points and the grey dashed line is the corresponding fit from \citet{AGH2017}.}
    \label{fig:sep}
\end{figure}

From our models, we obtain a similar scaling law in the asymptotic regime, that is
\begin{equation}
\frac{\overline{\rho}}{\overline{\rho}_\odot} = 
                   1.22\pm0.02 \left( \frac{\Delta\nu}{\Delta\nu_\odot} \right)^{2.091\pm 0.02}.
\label{eq:us}
\end{equation}
Note that the errorbars in equation~\ref{eq:agh} are observational, while those
of equation~\ref{eq:us} come from the fitting process.  While the coefficients
in our relation fall within the uncertainties of the observational relation, we
find a difference in the constant factor (corresponding to the offset between
the two trends in fig.~\ref{fig:sep}). 

\subsubsection{Effect of the metallicity}
The stars observed by \citet{AGH2017} have metallicities in the range $Z\sim
0.008 - 0.02$, which is covered by our models at $Z=0.01$ or $Z=0.02$.  As can
be seen on figure~\ref{fig:sep}, models at $Z=0.01$ tend to be slightly denser
than their $Z=0.02$ counterparts.  However, the large separation derived from
their island-mode spectrum follows closely the same scaling law: the
metallicity variations in the (narrow) range corresponding to the observations
has no impact on the $\overline{\rho} - \Delta\nu$ relation.

\subsubsection{Roche model}
Mean densities for eclipsing binary stars are derived by computing Roche model
surfaces, which rely on the simplifying assumptions that the stellar mass is
concentrated in its center and is rotating uniformly.  The stellar volume is
computed supposing, for simple geometrical reasons, that the radius measured
through the eclipse analysis is the equatorial one.  In order to test these
assumptions and their impact on the estimate of the volume, we compute the
volume of both the ESTER model and a Roche model of the same equatorial radius.

The volume of a given ESTER model is obtained through the integral
\begin{equation}
    V = \iiint r^2 \sin\theta d\theta d\phi dr = 
        4\pi \int^{\pi/2}_{0} \frac{R_s^3(\theta)}{3} \sin\theta d\theta,
\end{equation}
where $R_s$ is the surface radius computed consistently with the distribution
of matter inside the star.

\begin{figure}
    \centering
    \includegraphics[width=0.85\columnwidth]{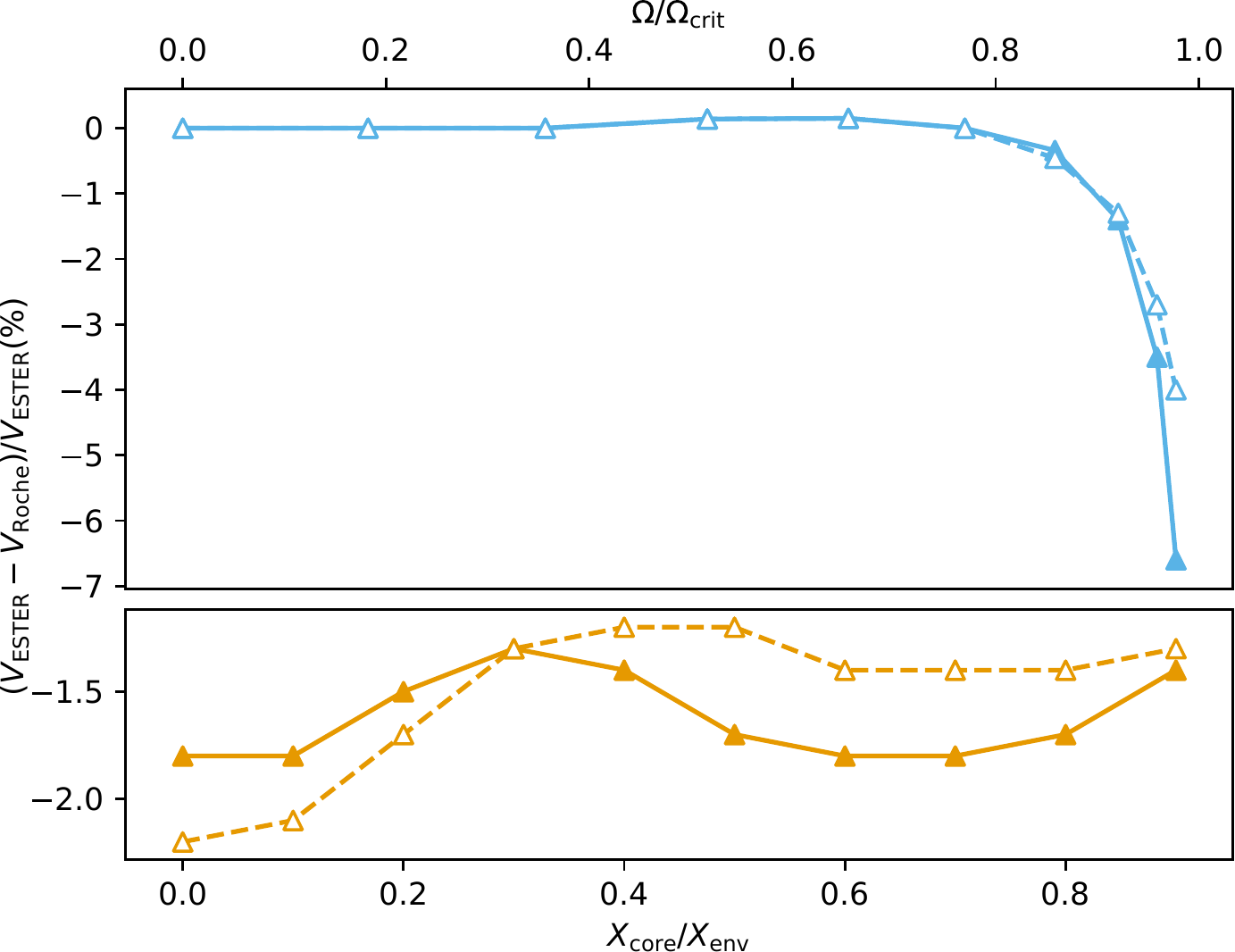}
    \caption{Relative difference in volume between Roche and ESTER models, in
    percentage, as a function of the rotation rate (top panel) or the core
    hydrogen abundance (bottom panel). The symbols are the same as in
    figure~\ref{fig:sep}.  The negative values show that Roche models tend to
    systematically underestimate the stellar volume, compared to ESTER models.
    }
    \label{fig:roche}
\end{figure}

Figure~\ref{fig:roche} shows the relative difference \mbox{$ \Delta{\rm
Volume}$}, defined as \mbox{$\left(\rm{V(Roche)-V(ESTER)}\right) / \rm
V(ESTER)$}.  We find a systematic difference: Roche models always underestimate
the volume, compared to models allowing for a more realistic distribution of
matter and rotation profile.  We see that the volume is underestimated by $\sim
1.6$\% on average, and does not seem to depend on the model core hydrogen
abudance. It remains low for all models, except at the most extreme rotation
rate (at 90\% of the critical velocity, where the difference can reach 6.6\%).
While this difference may impact the determination of stellar mean densities
and interferometric radii, for instance, it still is one order of magnitude too
small to account for the offset between the scaling relations
eq.~(\ref{eq:agh}) and eq.~(\ref{eq:us}). 

\subsubsection{Below the asymptotic regime}
We contend that the discrepancy between the scaling relation obtained through
modelling and that inferred from the observations lies with the frequency
regime in which we compute the large separation.  Indeed, in order to compare
our results with the predictions of \citet{LG09}, we computed high-frequency
modes so to place ourselves in the asymptotic regime, where the large
separation is expected to be constant.

However, \citet{AGH2009} showed that the frequency domain in which stars are
observed is far below the asymptotic domain.  They also showed that the large
frequency separation increases with frequency: the leftward drift of the ridges
shown in figure~\ref{fig:pattern} is a signature of this phenomenon.  Their
calculations show that computing the large separation in the asymptotic domain
can lead to a 10 to 15\% overestimate with respect to the observations. 

\begin{figure}
    \centering
    \includegraphics[width=\columnwidth]{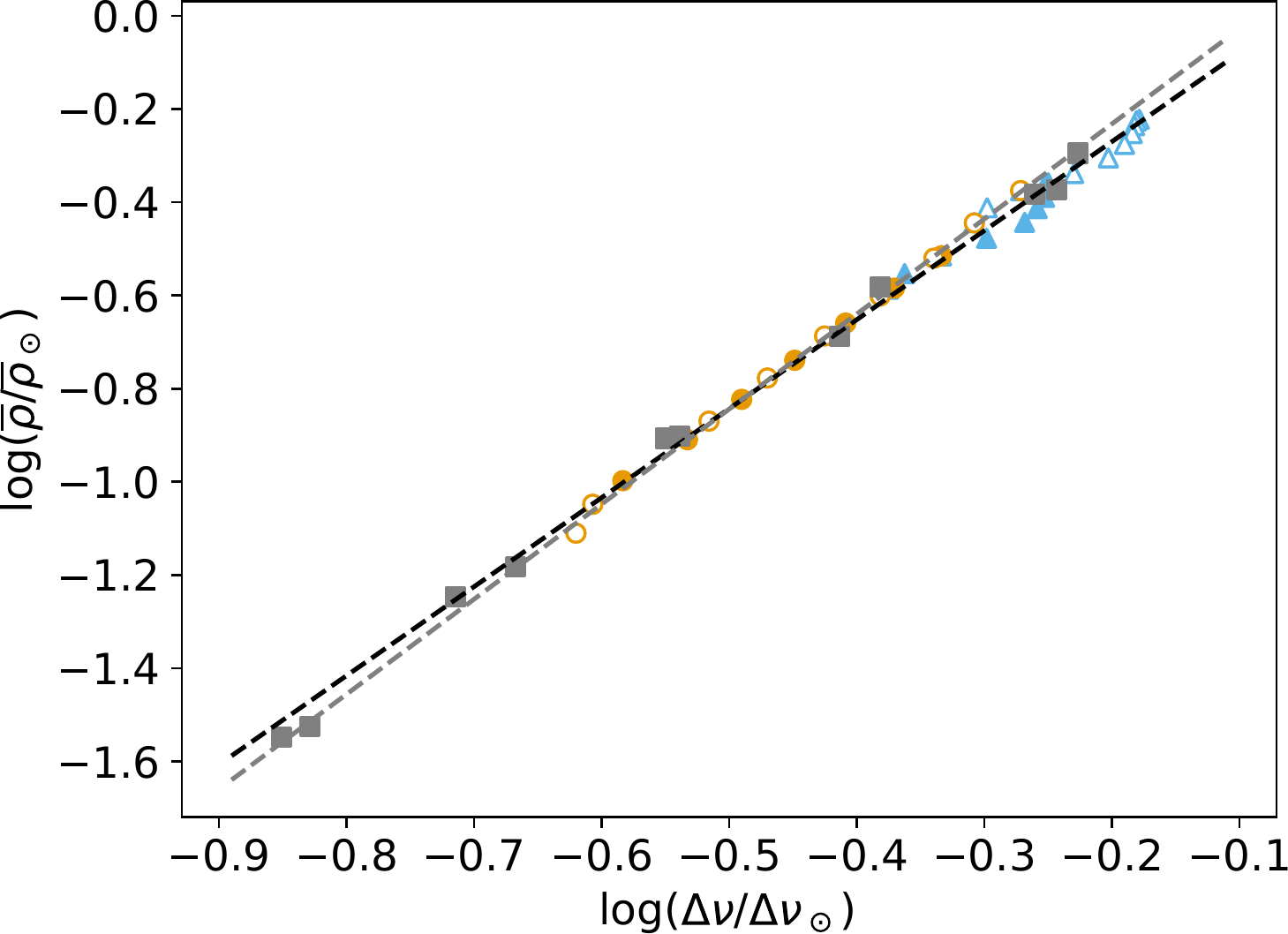}
    \caption{
    Stellar mean density as a function of the island-mode large separation in
    the frequency range corresponding to observations. Symbols are identical to
    those of figure~\ref{fig:sep}.
    }
    \label{fig:sep10percent}
\end{figure}

To investigate this explanation, we compute island modes at lower frequencies.
The stars observed by \citet{AGH2017} pulsate in different frequency ranges,
that we compare to the breakup rotation rate of each star: those pulsation
domains overlap in the range of $6$ to $9$ times the equatorial Keplerian
velocity. The island modes in this range have orders $\widetilde{n} = 8 - 12$,
these values are consistent with the range of excitable modes predicted by
\citet{Dupret05} (note that, for even $\tilde{\ell}=0, m=0$ modes,
$\tilde{n}=2n$).  In comparison, the asymptotic regime is reached at
frequencies roughly three times larger in our models, for $\tilde{n}= 32-40$. 

Figure~\ref{fig:sep10percent} shows the separation computed in this domain, and
qualitative agreement. The corresponding trend follows the relation 
\begin{equation}
\frac{\overline{\rho}}{\overline{\rho}_\odot} = 
                   1.30\pm0.06 \left( \frac{\Delta\nu}{\Delta\nu_\odot} \right)^{1.905\pm 0.03}.
\label{eq:us10}
\end{equation}
We find that the relation obtained using island modes fits the observations
better than that computed in the asymptotic regime. The small difference with
the observed relation may come from the reduced number of models we used.
Indeed, not all the models we computed were included, as numerical errors
prevented us from finding enough low-frequency island modes to compute a
reliable separation in the most evolved cases. A wider domain in densities
could be explored by computing models of different masses.

\begin{figure}
    \centering
    \includegraphics[width=0.85\columnwidth]{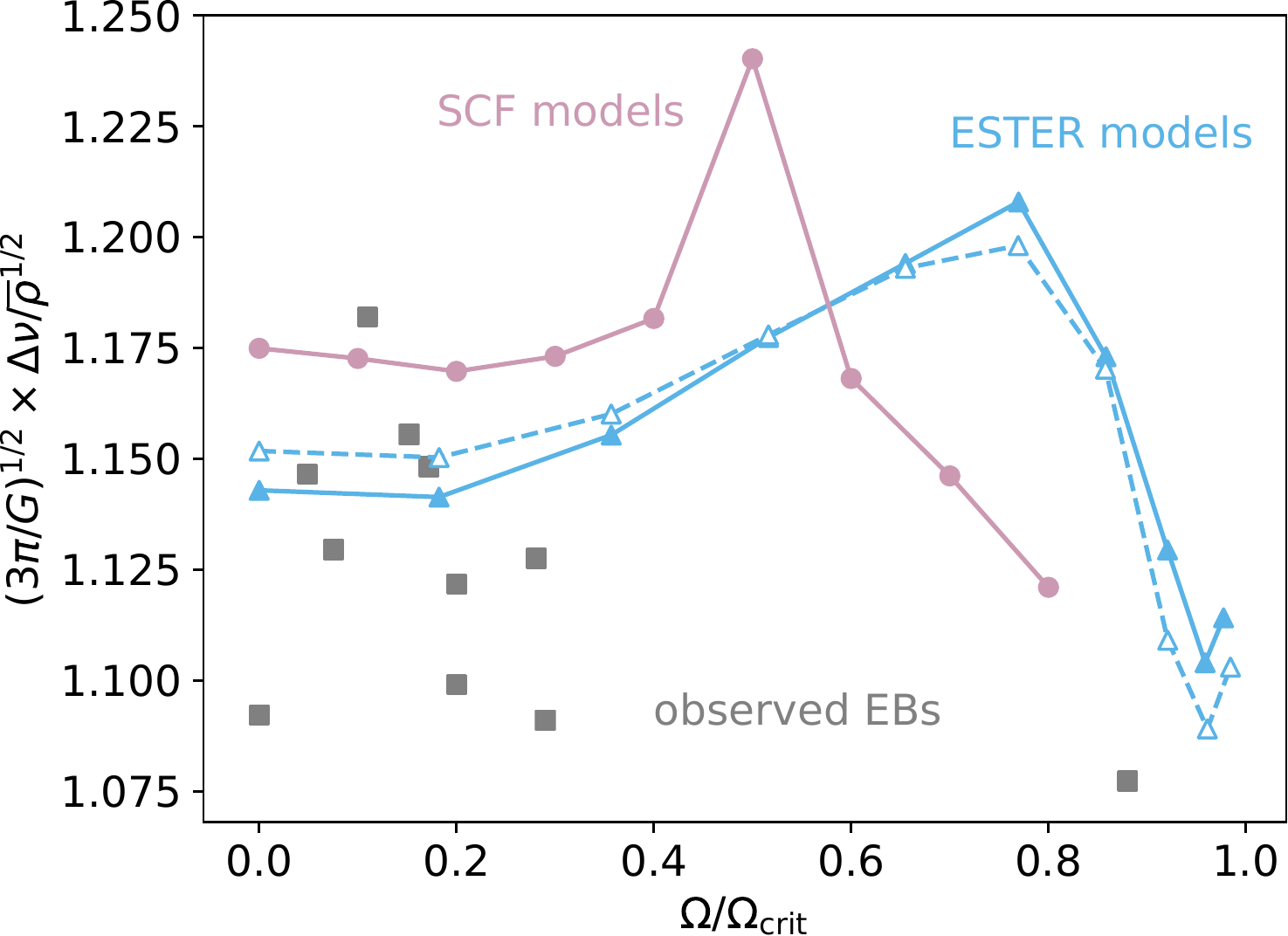}
    \caption{$\Delta\nu / \sqrt{\overline{\rho}}$ as a function of the rotation
    velocity (in units of the critical velocity). We use the values of
    $\Delta\nu$ obtained outside of the asymptotic (as in
    figure~\ref{fig:sep10percent}). We show the values obtained with $Z=0.02$
    and $Z=0.01$ ESTER models (full blue symbols and solid line, and empty blue
    symbols and dashed line, respectively), and compare them with SCF models
    from \citet{RTMJSM09} (purple, see main text) and the observations from
    \citet{AGH2017} (grey). The lack of a clear trend confirms that the scaling
    does not depend explicitly on rotation.  }
    \label{fig:rot}
\end{figure}

In Figure~\ref{fig:rot} we plot the ratio $ \Delta\nu/\sqrt(\overline{\rho})$
as a function of the rotation velocity. It shows the results obtained from
ESTER models, along with simpler $2 M_\odot$ SCF models~\citep{RTMJSM09} and
observational data~\citep{AGH2017}. In order to match the results for ESTER
models, the data points for SCF models have been recomputed in the frequency
domain corresponding to observations (the outlier at $\Omega = 0.5 \Omega_{\rm
crit}$ is due to an avoided crossing between some of the computed modes).  Some
differences can be expected between the results for ESTER and SCF models, given
that the former fully solve the structural and energy conservation equations,
thus resulting in a 2D rotation profile and a baroclinic structure, whereas the
latter only solves a horizontal average of the energy equation and uses imposed
cylindrical (or in our case, uniform) rotation profiles, thus resulting in a
simpler barotropic stellar structure. Nonetheless, the differences on
$\Delta\nu$ remain relatively small. The lack of a trend with rotation shows
that rotation has no additional impact on the large separation.

\section{Conclusions}
\subsection{Summary}
In this work, we computed two-dimensional models of fast-rotating stars and the
corresponding oscillations by interfacing the ESTER and TOP codes.  In order to
compare to observations and conduct scientific analysis, it was necessary to
first sort through the numerous modes calculated for every model.  We trained a
CNN to automate this process and classified the modes based on their geometry.
Our network architecture was designed to be versatile and allowed us to achieve
high accuracy whilst expediting the process dramatically.

In this first application of the deep-learning classification algorithm, we
focused on identifying a specific subclass of pressure modes present in fast
rotators, namely island modes of period 2.  Those modes are expected to be the
most visible in the p-mode frequency range, and to follow regular frequency
patterns in the high-frequency asymptotic regime.  We recover such patterns
with state-of-the-art models, confirming both previous theoretical and
observational works.  Previous work has linked the large frequency separation
of observed modes with the stellar mean density through a scaling law.  We find
a similar relation using the island-mode large separation, with an offset.  We
find that this difference cannot be attributed to metallicity effects nor to
the estimate of the stellar volume, but arises from `the difference in the
frequency domain sampled by the observed modes and the asymptotic domain in
which we study the synthetic oscillations.  Indeed, the modes observed in
actual stars are not in the asymptotic regime and therefore present large
separations roughly 10\% smaller: this difference can also be used to obtain an
estimate of the radial order of the detected oscillations in a given star.
This $\overline{\rho}-\Delta\nu$ relation obtained from the observations is a
very useful guide in modelling p-mode pulsators, such as $\delta$ Scuti stars,
and will in turn help mode identification and the matching between models and
observed stars.

\subsection{Future prospects}
We note that there is scope to optimize the CNN performance further, and we
will continue to do so in future work. We can also modify the CNN to use
quarter-plane plots, thus increasing the density of pixels by a factor of 2.
Such an improvement would require the creation of separate training sets for
the odd and even modes, which we are developing for future work.

Once the CNN had identified the 2-period island modes, they were manually
sorted according to their spherical degree $\widetilde{\ell}$. This subsequent
classification step can also be automated with a CNN and indeed the current
work has provided us with a substantial training set to do so.  We report an
accuracy of $>$99\% from our validation tests and have since added this
automated classification step in our analysis pipeline for future use.

Exploring a wider range of models, varying other parameters (and most notably
the stellar mass) will allow us to determine a more accurate and general
$\overline{\rho} - \Delta\nu$ relation.  There are other features of the
computed oscillation spectra that can be exploited, such as the separation of
modes at same $\tilde{n}$ and consecutive $\widetilde{\ell}$ values.  Varying
the azimuthal order $m$ will also allow us to bring out rotational splittings
(that is, the frequency separation between modes with the same
$(\widetilde{\ell},\widetilde{n})$ values but different $m$ values, which carry
the signature of the stellar rotation), allowing for a description of the
internal (differential) rotation of the star (Reese et al. in prep).  The last
remaining step is the automatic determination of the radial order $\tilde{n}$,
which would allow the derivation of accurate asymptotic formulae for fast
rotating stars.  We note that the conclusions of this work have to be linked
with previous efforts towards mode identification in fast-rotating stars, such
as the calculation of mode visibilities \citep{R2013} or two-dimensional
non-adiabatic pulsation computations \citep{MRRB17}.  Finally, expanding the
number of stars on which this technique can be applied through current and
future asteroseismology missions such as BRITE, TESS, or PLATO,  will be of
great help to confirm and elucidate any hidden dependence in the obtained
scaling relation. 

\section*{Acknowledgements}
The authors thank Dr. Marc-Antoine Dupret for useful comments that helped
improve this letter considerably. We also thank Dr. Antonio Garc{\'\i}a
Hern{\'a}ndez for fruitful discussions and Dr. James Kuszlewicz for providing
the training data for the Kepler giants.  GMM and DRR benefited from the
hospitality of ISSI as part of the SoFAR team in early 2018.  DRR acknowledges
the support of the French Agence Nationale de la Recherche to the ESRR project
under grant ANR-16-CE31-0007 as well as financial support from the Programme
National de Physique Stellaire of the CNRS/INSU co-funded by the CEA and the
CNES.



\bibliographystyle{mnras}

\label{lastpage}
\end{document}